\documentclass[conference,letterpaper]{sig-alternate-10pt}
\usepackage{amssymb}
\usepackage{bbm}
\usepackage{tikz}
\usepackage{subfigure}
\usetikzlibrary{arrows,automata}
\usepackage{amsmath}
\usepackage{empheq}
\usepackage{lastpage}
\usepackage{graphicx}
\usepackage{color}
\usepackage{multirow}
\usepackage[hang,scriptsize,bf,margin=0cm]{caption}

\begin{document}

\conferenceinfo{}{}
\title{StreaMon: a data-plane programming abstraction for Software-defined Stream Monitoring\titlenote{\tiny This work will be submitted for possible
publication. Copyright may be transferred without notice, after
which this version may no longer be accessible.}}

\numberofauthors{1}
\author{Giuseppe Bianchi, Marco Bonola, Giulio Picierro, Salvatore Pontarelli, Marco Monaci\\
	\affaddr{University of Rome "Tor Vergata"}\\
	\email{name.surname@uniroma2.it}
}

\maketitle

\begin{abstract}
The fast evolving nature of modern cyber threats and network monitoring needs calls for new, ``software-defined'', approaches to simplify and quicken programming and deployment of online (stream-based) traffic analysis functions. StreaMon is a carefully designed data-plane abstraction devised to scalably decouple the ``programming logic'' of a traffic analysis application (tracked states, features, anomaly conditions, etc.) from elementary primitives (counting and metering, matching, events generation, etc), efficiently pre-implemented in the probes, and used as common instruction set for supporting the desired logic. Multi-stage multi-step real-time tracking and detection algorithms are supported via the ability to deploy custom states, relevant state transitions, and associated monitoring actions and triggering conditions. Such a separation entails platform-independent, portable, online traffic analysis tasks written in a high level language, without requiring developers to access the monitoring device internals and program their custom monitoring logic via low level compiled languages (e.g., C, assembly, VHDL). We validate our design by developing a prototype and a set of simple (but functionally demanding) use-case applications and by testing them over real traffic traces.  
\end{abstract}


\section{Introduction}
\label{s:intro}
The sheer volume of networked information, in conjunction 
with the complexity and polymorphous nature of modern cyberthreats, 
calls for scalable, accurate, and, {\em at the same time}, flexible 
and programmable monitoring systems \cite{deri10, dain12}. 
The challenge is to {\em promptly} react to fastly mutating needs by 
deploying {\em custom traffic analyses}, capable of tracking event chains 
and multi-stage attacks, and efficiently handle the many heterogeneous 
features, events, and conditions which characterize an operational failure, 
a network application mis-behavior, an anomaly or an incoming attack. 
Such needed level of flexibility and programmability should address 
scalability {\em by design}, through systematic exploitation of 
stream-based analysis techniques. And, even more challenging, traffic 
analyses and mitigation primitives should be ideally brought 
{\em inside} the monitoring probes themselves at data plane, so as to avoid 
exporting traffic data to central analysis points, an hardly adequate way 
to cope with the traffic scale and the strict (ideally real time) mitigation 
delay requirements. 

\begin{figure*}[t]
\centering
   \includegraphics[width=17cm, height=3cm]{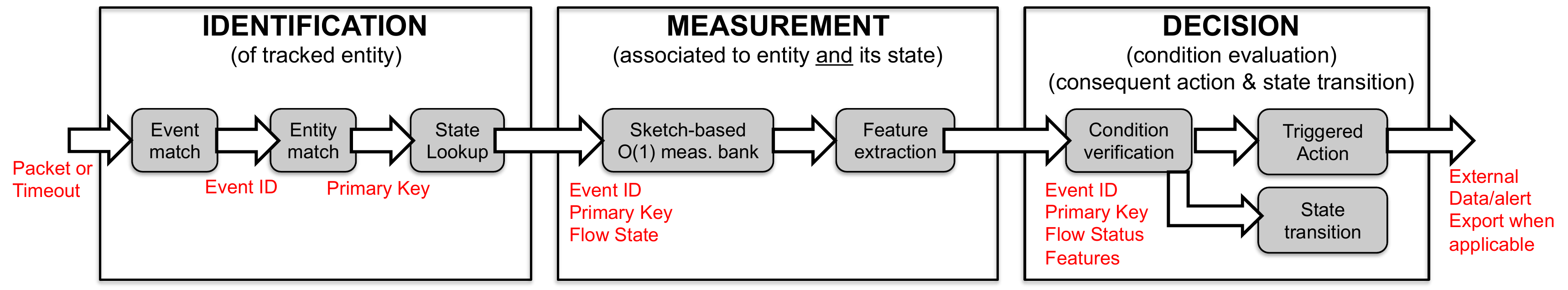}
\caption{StreaMon data plane identification/measurement/decision abstraction, and its mapping to implementation-specific workflow tasks}
\vspace*{-.5cm}
\label{fig:logic-wf}
\end{figure*}

To face this rapidly evolving scenario, in this paper we propose StreaMon, a data-plane 
programming abstraction for stream-based monitoring tasks directly running over network probes. 
StreaMon is devised as a pragmatic tradeoff between full programmability and 
vendors' need to keep their platforms {\em closed}. StreaMon's strategy  
closely resembles that pioneered by Openflow \cite{OF08} in the abstraction 
of networking functionalities, thus paving the road towards software-defined networking\footnote{
	And in fact, even if the focus of this paper is (as a first step) only on the data plane 
	interface itself, we believe that StreaMon could eventually candidate to play the 
	role of a possible southbound data-plane interface for Software-Defined Monitoring frameworks, 
	devised to translate high level monitoring tasks into a sequence of low-level 
	traffic analysis programs distributed into network-wide StreaMon probes.}. 
However, the analogy with Openflow limits to the strategic level; in its technical design, 
StreaMon significantly departs from Openflow for the very simple reason that (as 
discussed in Section \ref{s:abstraction}) the data-plane programmability of monitoring tasks 
exhibits very different requirements with respect to the data-plane programmability of networking 
functionalities, and thus mandate for different programming abstractions. \\
The actual contribution of this paper is threefold. 

{\bf (1)} We identify (and design an execution platform for) an extremely simple abstraction which appears capable of supporting a wide range of monitoring application requirements. The proposed API decouples the monitoring application ``logic'', externally provided by a third party programmer via (easy to code) eXtended Finite State Machines (XFSM), from the actual ``primitives'', namely configurable sketch-based measurement modules, d-left hash tables, state management primitives, and export/mitigation actions, hard-coded in the device. While our handling of sketch-based measurement primitives may recall \cite{opensk}, we radically differentiate from any other work we are aware of, in our proposal to inject monitoring logic in the form of XFSM devised to locally orchestrate and run-time adapt measurements to the tracked state for each monitored entity (flows, hosts, DNS names, etc),
with no need to resort to an external controlling device.

{\bf (2)} We implement two StreaMon platform prototypes, a full SW and a FPGA/SW integrated implementation. We functionally validate them with five use case examples (P2P traffic classification, Conficker botnet detection, packet entropy HW analysis, DDos detection, Port Knocking), not meant as stand-alone contributions, but rather selected to showcase the StreaMon's adaptability to different application requirements. 

{\bf (3)} We assess the performance of the proposed approach: even if the current prototype implementation is not primarily designed with performance requirements in mind, we show that it can already sustain traffic in the multi-gbps range even with several instantiated metrics (for example 2.315 Gbps of real world replayed traffic with 16 metrics, see section \ref{s:perf}); moreover, we show that scalability can be easily enhanced by offloading the SW implementation with HW accelerated metrics; in essence, it seems fair to say that scalability appears to be an {\em architectural} property of our proposed API, rather than a side effect of an efficient implementation.


\section{StreaMon abstraction}
\label{s:abstraction}

Our strategy in devising an abstraction for deploying stream-based monitoring tasks over a (general-purpose) network monitoring probe is similar in spirit to that brought about by the designers of the Openflow \cite{OF08} match/action abstraction, for programming networking functionalities over a switching fabric. Indeed, we also aim at identifying a compromise between full programming flexibility, so as to adapt to the very diverse needs of monitoring application developers and permit a broad range of innovation, and consistency with the vendors' need for closed platforms. However, the requirements of monitoring applications appear largely different from that of a networking functionality, and this naturally drives towards a {\em different} pragmatic abstraction with respect to a match/action table. 

At least three important differences do emerge. First, the ``entity'' being monitored is not consistently associated with the same field (or set of fields) in the packet header. For instance, if the target is to detect whether an IP address (monitored entity) is a bot, and the chosen mechanism is to analyze if the percentage of DNS NXDomain replies (feature) is greater than a given threshold (condition), the flow key to use for accounting is the source IP address when the arriving packet is a DNS query (event), but becomes the destination IP address when the packet is a DNS response (a different event). 

Second, the type of analysis (and possibly the monitoring entity target) entailed by a monitoring application may change over time, dynamically adapting to the knowledge gathered so far. For instance, if an IP address exhibits a critical percentage of DNS NXDomain replies, hence a bot suspect, we may further track its TCP SYNACK/SYN ratio and determine whether horizontal network scans occur, so as to reinforce our suspicion. And we might then follow up by deriving even more in-depth features, e.g., based on deep packet inspection. But at the same time, we would like to avoid tracking {\em all} features for {\em all} the possible flows, as this would unnecessarily drain computational resources. 

Finally, activities associated to a monitoring task are not all associated to a matching functionality: only measurement and accounting tasks are. Rather, less frequent, but crucial, activities (such as forging and exporting alerts, changing states, setting a mitigation filtering rule, and so on) are based on {\em decisions} taken on what we learned so far, and which are hence triggered by {\em conditions} applied to the gathered features. 

Our proposed StreaMon abstraction, illustrated in figure \ref{fig:logic-wf}, appears capable to cope with such requirements (as more extensively shown with the use cases presented in section \ref{ss:usecases}). It comprises of three ``stages'', programmable by the monitoring application developer via external means (i.e. not accessing the internal probe platform implementation).

\begin{figure}[!t]
\centering
   \includegraphics[width=.45\textwidth]{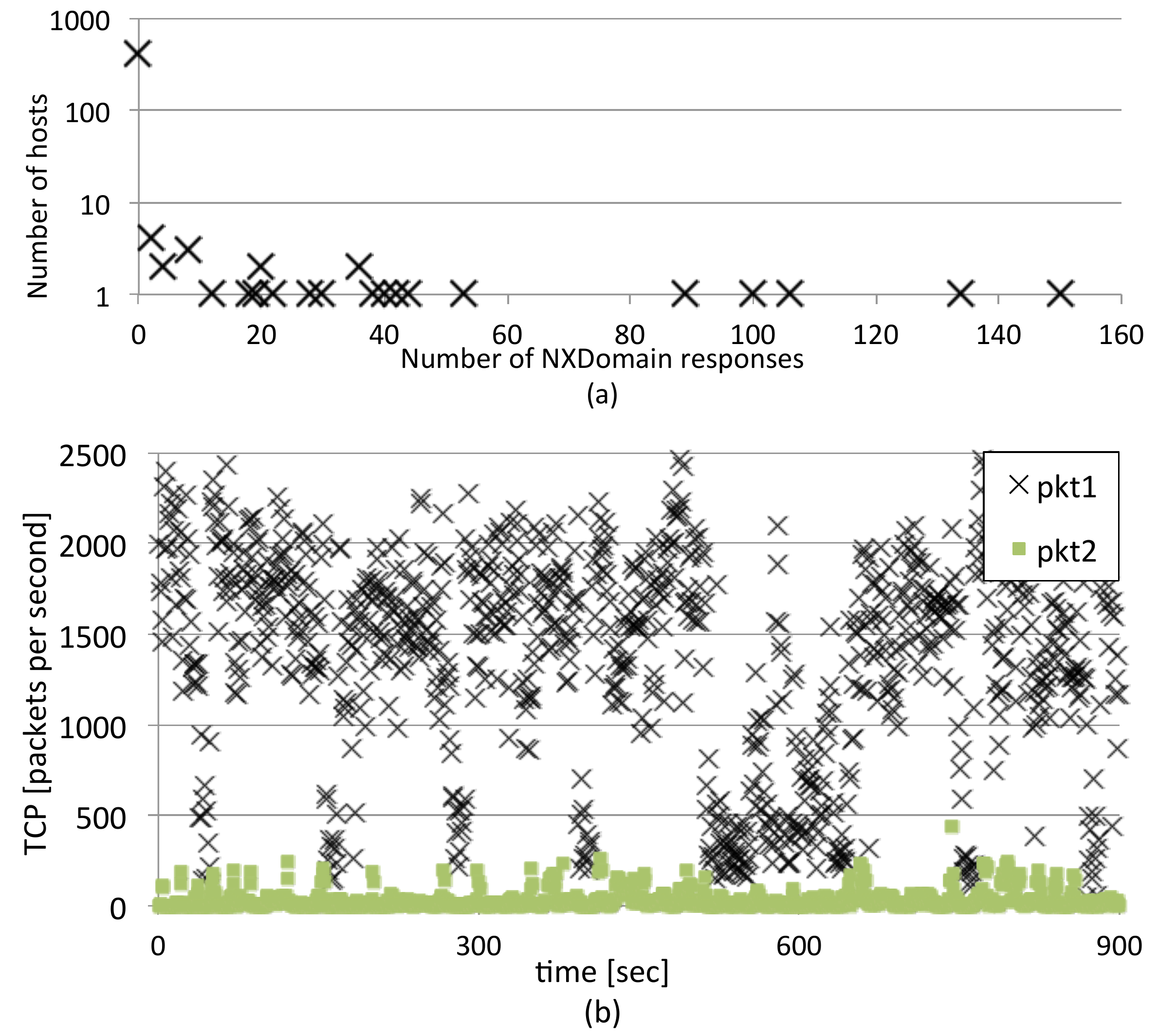}
\caption{Experimental analysis of number of NXDomain DNS responses received (a feature commonly used in Botnet detection, e.g., \cite{confickerc}). Data obtained using a trace of 155 minutes with 955 different hosts belonging to 36 different /24 subnetworks. (a): Number $F_1$ of DNS NXDomain responses received per host (430 IPs performing at least 1 DNS query, 401 of them don't receive any NXDomain response); (b): number of tcp packet per seconds for all the flows (pkt1) and only for the flows for which $F_1 >= 1$ (pkt2), 15 minutes sample; analysis time for whole pkt1 trace: 1179 seconds, versus 68 seconds for pkt2.}
\vspace*{-.5cm}
\label{fig:offloading}
\end{figure}

{\bf (1)} The {\bf Identification} stage permits the programmer to specify what is the monitored entity (more precisely, deriving its primary key, i.e., a combination of packet fields that identify the entity) associated to an event triggered by the actual packet under arrival, as well as retrieve an eventually associated state\footnote{
	Our implementation uses d-left hashes for O(1) complexity in managing states. Also, memory consumption 
	can be greatly reduced by {\em opportunistically} storing only non-default (e.g. anomalous) states. See the 
	illustrative experiment in Fig. \ref{fig:offloading} for a rough idea of the attainable saving in the Conficker 
	detection example, use case \ref{ss:use-conficker}.}.

{\bf (2)} The {\bf Measurement} stage permits the programmer to configure which information should be accounted. It integrates {\em hard-coded and efficiently implemented} hash-based measurement primitives (metric modules), fed by configurable packet/protocol fields, with externally programmed {\em features}, expressed as arbitrary arithmetic operations on the metric modules' output. 

{\bf (3)} The {\bf Decision} stage is the most novel aspect of our abstraction. It permits to take decisions programmed in the form of eXtended Finite State Machines (XFSM), i.e. check conditions associated to the current state and tested over the currently computed features, and trigger associated actions and/or state transitions. 

The obvious compromise in our proposed approach is that (as per the match/action Openflow primitives) new metrics or actions can only be added by directly implementing them over the monitoring devices, thus extending the device capabilities. However, even restricting to our actual StreaMon prototype, the subset of metrics we implemented appear reusable and sufficient to support features proposed in a meaningful set of literature applications - see Table \ref{tab:features} in Section \ref{s:metrics-features}.


\section{StreaMon Processing Engine}
\label{s:at-a-glance}

\subsection{System components}
\begin{figure}[!tb]
\centering
   \includegraphics[width=.5\textwidth]{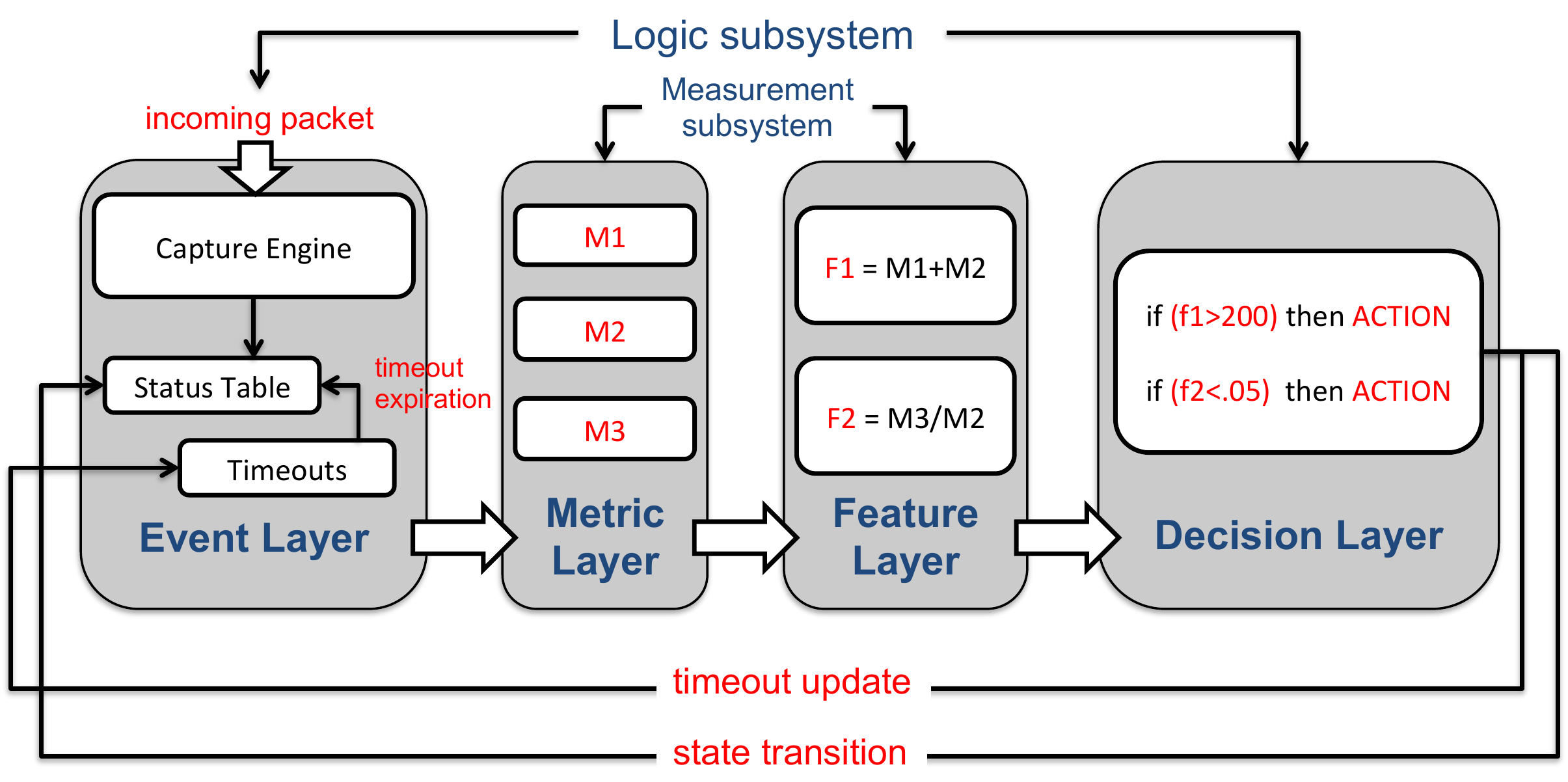}
\caption{StreaMon processing engine architecture}
\label{fig:arch_overview}
\vspace{-.5cm}
\end{figure}

The previously introduced abstraction can be concretely implemented by a stream processing engine whose architecture is depicted in Figure \ref{fig:arch_overview}. It consists of four modular layers descriptively organized into two subsystems, namely {\em Measurement subsystem} and {\em Logic subsystem}, detailed in sections \ref{s:metrics-features} and \ref{s:tracking}, respectively.

\textbf{Event layer} - Such layer is in charge of parsing each raw captured packet, and match an {\em event} among those user-programmed via the StreaMon API. The matched event identifies a user-programmed \textit{primary key} which permits to retrieve an {\em eventually} stored state. The event layer is further in charge of supplementary technical tasks (see section \ref{s:tracking}), such as handling special timeout events, deriving further secondary keys, etc.

\textbf{Metric layer} - StreaMon operates on a per-packet basis and does {\em not} store any (raw) traffic in a local database. The application programmer can instantiate a number of {\em metrics} derived by a basic common structure, implemented as computation/memory efficient multi-hash data structures (i.e., Bloom-type sketches), updated at every packet arrival. 

\textbf{Feature layer} - this layer permits to compute user-defined arithmetic functions over (one or more) metric outputs. Whereas metrics carry out the bulky task of accounting {\em basic} statistics in a scalable and computation/memory efficient manner, the features compute {\em derived} statistics tailored to the specific application needs, at no (noticeable) extra computational/memory cost.

\textbf{Decision layer} - this final processing stage implements the actual application logic. This layer keeps a list of \textit{conditions} expressed as mathematical/logical functions of the feature vector provided by the previous layer and any other possible secondary status. Each condition will trigger a set of specified and pre-implemented \textit{actions} and a state \textit{transition}.

\begin{figure}[t]
	\centering
	\includegraphics[width=.5\textwidth]{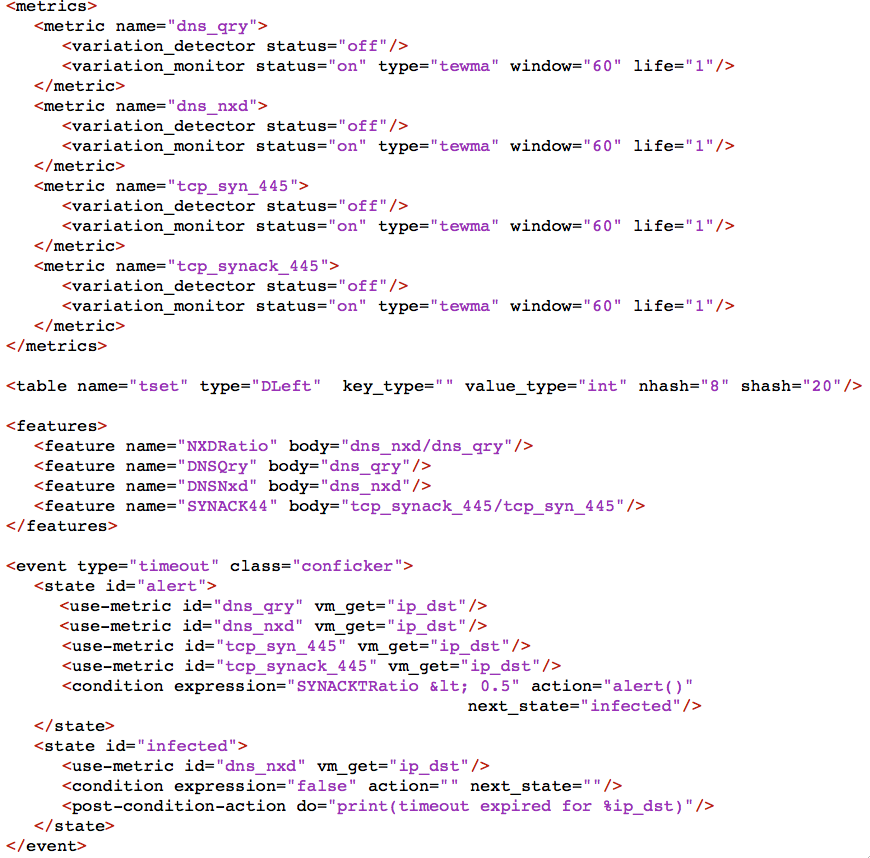}
	\caption{Excerpt of XML based StreaMon code showing: (i) metric element allocation, (ii) feature compositions, (iii) an event logic description. In particular this picture shows a timeout event handler, described in terms of metric operations, feature extractions, conditions, actions and state transition}
	\label{fig:xml}
\end{figure}

\subsubsection*{StreaMon's programming language} 
Application programmers describe their desired monitoring operations 
through an high-level XML-like language, which permits to specify custom (dynamic) states, configure 
measurement metrics, formalize when (e.g.in which state and for which event) and how (i.e. by 
performing which operations over available metrics and state information) to extract 
features, and under which conditions trigger relevant actions (e.g. send an alert or data 
to a central controller). We remark that a monitoring application formally specified 
using our XML description does not require to be {\em compiled} by application developers, 
but is run-time installed, thus significantly simplifying on-field deployment. Figure \ref{fig:xml} shows an excerpt of a StreaMon application code.

\subsubsection*{HW acceleration}
StreaMon allows the seamless integration of HW accelerated metrics, e.g. mandated by stringent performance requirements. This remains transparent to the application programmer, which can thus port the same application from a SW based probe to a dedicated HW platform {\em with no changes in the application program}. Seamless HW integration is technically accomplished by performing {\em all} metrics, parsing, and event matching HW-accelerated computations in a front-end (in our specific case, an FPGA), and by bringing the relevant results up to the user plane through a HW/SW interface,  by appending the meta-data generated by the HW tasks to the packet.


\subsection{Measurement Subsystem}
\label{s:metrics-features}
The StreaMon Measurement subsystem provides a fast, computation/memory efficient, set of n highly configurable built-in modules that allows a programmer to deploy and compute a wide range of traffic features in stream mode. 

\begin{figure}[!t]
\centering
\includegraphics[width=.45\textwidth]{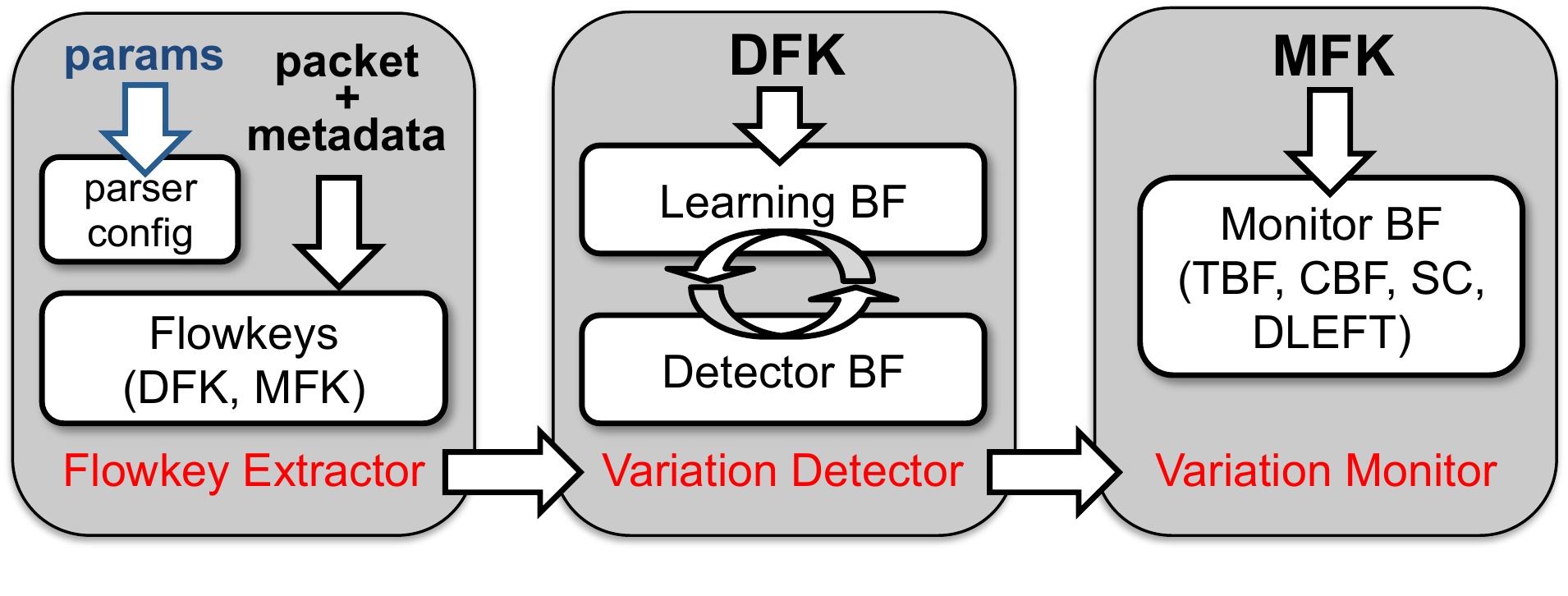}
\vspace*{-.5cm}
\caption{Metric module structure}
\vspace*{-.5cm}
\label{fig:bf-metric}
\end{figure}

\textbf{Multi-Hash Metric module (MH)} - From a high level point of view, StreaMon metric modules are functional blocks exporting a simple interface to update and retrieve metric values associated to a given key. Even though in principle such modules can be implemented with any compact data structure compatible to our proposed stream mode approach (and indeed the current architecture support pluggable user defined metric modules),  several metrics of practical interest can be derived from a basic structure depicted in Figure \ref{fig:bf-metric} and extending the construction proposed in \cite{info10Bianchi}. It permits to count (or time-average, see below) {\em distinct} values (called variations in \cite{info10Bianchi}) associated to a same key; for example: the number of {\em distinct} IP destinations contacted by a same IP source, or the number of distinct DNS names associated to a same IP address. The MH module is implemented using Bloom filter extensions \cite{broder02, info10Bianchi, ccr11Bianchi}. Processing occurs in three stages: (i) \emph{extract} from the packet\footnote{
	Even if, for simplicity of explanation, we account such an extraction 
	to the MH module, for obvious performance reasons we perform packet 
	parsing and the consequent extraction of the flowkeys DFK and MFK once 
	for all, for all metrics, in the Event Layer. Indeed, different metrics may make
	usage of a same packet field.}
two binary strings, \emph{Detector FlowKey} (DFK) and  \emph{Monitor FlowKey} (MFK), that will be used in to query the subsequent filters; (ii) \emph{detect} whether the DFK has already appeared in a past time window (\emph{Variation Detector filter}, VD), and if this is the case, iii) \emph{account} to the MFK key some packet-related quantity (e.g. number of bytes, or simply add 1 if packet count is targeted) to the third stage's  \emph{Variation Monitor} (VM) filter. The reader interested in the rationale behind such construction and in further design details may refer to \cite{info10Bianchi}. Finally note that, as a special case, the MH module can be configured to either i) provide an ordinary count, by disabling the VD stage, or ii) perform a match for checking whether some information associated to the incoming packet is first seen (in a given time window), by disabling the VM stage.

\begin{table*}[t]
\hfill{}
\begin{footnotesize}
\begin{tabular}{| l | p{3.7cm}| p{11.8cm}|}
\hline
\textbf{Paper}&\textbf{Description}&\textbf{Brief description of reference features supported by StreaMon}\\
\hline
\cite{exposure}&Passive DNS anal. x malicious domain detection&(i) Short life and Access Ratio per domain; (ii) Multi-homing/multi-address per domain; (iii) AVG, STD, total number, CDF of the TTL for a given domain\\
\hline
\cite{complex}&Traffic charact. via Complex Network modeling&CDF, STD, Max/Min of: (i) total number of endpoints and correspondent host per endpoint; (ii) bytes exchanged between two endpoints\\
\hline
\cite{fastflux}&DNS analysis for fastflux domain detection&\textit{Arithmetic functions of:} (i) number of unique A records returned in all DNS lookups, (ii) number of A records in a DNS response, (iii) number of name server (NS) records in one single lookup, (iv) number of unique ASNs for each A record\\
\hline
\cite{scidive}&Stateful VOIP attack detection&\textit{Stateful rules:} (i) RTP traffic after SIP BYE? (ii)  source IP address changed in time window? (iii) RTP sequence numbers evolve correctly?\\
\hline
\cite{ddos02}&DDOS detection through ICMP and L4 analysis&Sent/received TCP packet ratio; Sent/received for different ICMP messages; UDP bitrate; No. different UDP connections per destination; Connection packet count\\
\hline
\cite{p2pSVM}&SVM-based method for P2P traffic classification&\textit{Per distinct host and port:} (i) ratio between TCP and UDP traffic; (ii) AVG traffic speed; (iii) ratio between TCP and UDP AVG packet length; \textit{Per distinct IP:} (i) ARP request count, (ii) ratio between sent/received {TCP, UDP} packet, (iii) ratio between \{TCP, UDP\} and total traffic; \textit{Per distinct port:} traffic duration\\
\hline
\cite{snare}&Mail spammer detection via Network level analysis& (i) AVG/STD msg len in past 24h; (ii) AVG distance to 20 nearest IP neighbors of the sender; (iii) AVG/STD of geodesic distance between sender and recipient; (iv) Message body len; (v) AVG, STD and total no. different recipients\\
\hline
\end{tabular}
\end{footnotesize}
\caption{}
\label{tab:features}
\end{table*}

\textbf{Programming the Measurement subsystem} - StreaMon programmers can deploy two types of ``counting'' data structures: (i) an ordinary \emph{Count Sketch} that sums a value associated to the MFK (CBF) or (ii) a \emph{Time decaying Bloom Filter} (TBF) \cite{ccr11Bianchi} that performs a smoothed (exponentially decaying) average of a value associated to the MFK. At {\em deployment time}, an MH module is configured by: (i) enabling/disabling the VD or the VM; (ii) specifying the counting data structure in the VM; (iii) setting the MH parameters, i.e., number of hash functions, total memory allocated, swapping threshold \cite{info10Bianchi} or memory time window in the VD, TBF's smoothing parameter \cite{ccr11Bianchi}; (iv) chain multiple MH module (so as to update a metric only if a former one was updated). At {\em run time}, the programmer may dynamically change flowkeys and updating quantities.  In particular, for each possible event and (if needed) flow status (i.e., XFSM entry), a list of Metric Operations (MOP) is defined, where a MOP is a set/get primitive that defines the flowkeys (as a packet fields combinations) and quantities to be monitored/accounted. Finally, a set of StreaMon metrics are combined into Features by simply defining mathematical functions $F_i=f_i(\overline{M})$, where $\overline{M}$ is a Metric vector.

\textbf{What practical features can be supported?} - We believe that the above metrics fulfill the needs of a non negligible fraction of real world monitoring applications. Indeed, \cite{ccr11Bianchi} shows a concrete example of a real world application reimplemented by just using modules derived from our general MH module described above. Indeed, limiting to the StreaMon's measurement subsystem (we will significantly extend the framework in the next section), table \ref{tab:features} shows features considered in a number of works taken from a somewhat heterogeneous set (different targets, operating at different network layers, different classification approaches) which, according to our analysis, are readily deployed through suitable configuration of the above metrics/features. Indeed, most applications either require to track/match (1) features that are directly retrievable from a single packet and do not have memory of past values (e.g.: short life of a DNS domain, message body length, all the features in \cite{fastflux} except the first one); and/or (2) require counting or averaging over a various set of parameters (eventually uniquely accounted - e.g. variations), which are readily instantiated with an MH module\footnote{
	For instance, the {\em number of bytes exchanged between two hosts} can be obtained by 
	setting as VMK is the concatenation of source and destination IP address and as updating 
	quantity the length field of the packet; similarly, unique counts such as the 
	\emph{total number of distinct TTL for a given DNS domain} is obtained by setting as VDK the concatenation 
	of Domain Name and TTL in the DNS response and as VMK is the Domain Name.};
and/or (3) require logical/mathematical combinations of different statistics, which is the goal of our Feature Layer.  Traffic features not covered by the families listed above are those which require a stateful correlation of different flows status. Such metrics (like the ones exploited in \cite{scidive}) are supported by StreaMon but require the stateful framework described in Section \ref{s:tracking}.


\subsection{Logic subsystem}
\label{s:tracking}
StreaMon's \emph{Logic Subsystem} is the result of the interwork between the {\em Event Layer} and the {\em Decision Layer}. It provides the application developer with the ability to \emph{customize the tracking logic} associated to a monitored entity {\em subject to specific user-defined conditions}, so as to provide a verdict and/or perform analyses beyond those provided by the Monitoring Subsystem. 

\textbf{Event Layer} - The \emph{Event Layer} generates the events triggering the StreaMon processing chain, identifies the monitored entity (primary key), and retrieves the specific event context (in particular the flow status). This layer is further composed of three functional component: (i) the \textit{capture engine} responsible for ``capturing'' the triggering event, either a ``packet arrival'' or a ``timeout expiration'';(ii) the \textit{timeout manager} in charge of keeping track of all registered timeouts and manage their expiration; (iii) the \textit{status table}, a (dleft hash) table storing the status associated to the processed flows - the primary key status - and the so-called \textit{secondary support table}, a table storing states not associated to the primary key but required by the application state machine to take some decision. 

The triggering event is associated to a \textit{primary key}, i.e. the monitored entity (flow, host, etc) under investigation for which the monitoring application will provide a final verdict, like ``infected'' or ``legitimate''. For intercepted packets, the primary key is a combination of packet fields parsed by the (extensible) protocol dissector implemented in this layer. For locally generated timeout expiration events, the primary key is retrieved from the timeout context. The primary key is used to retrieve the current status of the flow: no match in the state table is considered to be a \emph{default} state.  

If the primary key is not directly retrievable from the packet, and instead the flow status is related to some other flow previously processed, a secondary support table storing the references to the entries in the status table is used. Such secondary support table is called \emph{related status table}. For example, this can be the case of an application keeping track of SIP URIs, (i.e.: the primary key is the SIP user ID in SIP INVITE messages) and consider as suspicious all UDP packets related to a data connection with a suspicious SIP user. In case of UDP packets, neither of the packet fields can be used to retrieve the flow status. Instead, a secondary support table is used to keep a reference between the socket 5-tuple and the status entry of the associated SIP initiator user. If the application does not take into account the flow status, the primary extraction is skipped.


\textbf{Decision Layer} - The Decision Layer is the last processing unit of the StreaMon architecture and receives the event context carrying an indication of the current flow status and a set of traffic features container. This layer keeps a list of \emph{Decision Entries} (DEs) defined as the 3-tuple \emph{(enabling Condition (C), Output actions (O), State Transition (ST))}. For each triggering event, and according to the current flow status, the decision layer verifies the enabling conditions and executes the actions and the state transition associated to the matched condition. Since secondary support tables can be updated, StreaMon support \emph{variable conditions}, i.e. in which the comparison operands may change during time. 
The first matched condition will trigger the execution of a action set (like DROP, ALERT, SET\_TIMEOUT etc..).

\textbf{Programming the Logic Subsystem} - Programmers describe an XFSM specifying (i) states and triggering events; (ii) for each state, which metrics and features are updated and which auxiliary information is collected/processed (secondary table); (iii) which conditions trigger a state transition and the associated actions. 

\begin{table}
\begin{center}
\begin{footnotesize}
\begin{tabular}{ | l | l | }
\hline
\multicolumn{2}{ |c| }{\textbf{Logic Subsystem built-in primitives}}\\
\hline
\multirow{3}{*}{\textbf{Operators}} & \texttt{SUM, DIFF, DIV, MULT, MOD} \\
& \texttt{EQ, NEQ, LT, GT}, \texttt{SQRT, LOG, POW} \\
& \texttt{AND, OR, NOT, XOR} \\
\hline
\multirow{3}{*}{\textbf{Actions}} & \texttt{SET\_TIMEOUT}, \texttt{UPDATE\_TIMEOUT}\\
& \texttt{SAVE\_TIMEOUT\_CTX}, \texttt{DROP}, \texttt{ALLOW}, \texttt{MARK} \\
& \texttt{NEXT\_STATUS}, \texttt{UPDATE\_TABLE}, \texttt{PRINT}, \texttt{EXPORT}\\
\hline
\end{tabular}
\end{footnotesize}
\end{center}
\vspace*{-.5cm}
\caption{}
\vspace*{-.5cm}
\label{tab:prim}
\end{table}

To identify the triggering event, StreaMon keeps a list of user-defined ``event descriptors'', expressed as a 3-tuple \textit{(type, descriptor, primary key)}. For example, in case of intercepted packets, an event can be defined as: $(packet; (udp.dport==53); ip.src)$.

Moreover, programmers are given primitives to define for each event and state: (i) a set of metric operations (MOP) as described in section \ref{s:metrics-features}; (ii) a set of conditions expressed as an arithmetic function of features and secondary support table values. For each condition, programmers define a set of built-in actions and a state transition. Table \ref{tab:prim} summarizes the logic subsystem supported condition operators and actions.

\section{Simple Use case examples}
\label{ss:usecases} 

\begin{figure*}[t]
	\centering
	\includegraphics[width=1.02\textwidth]{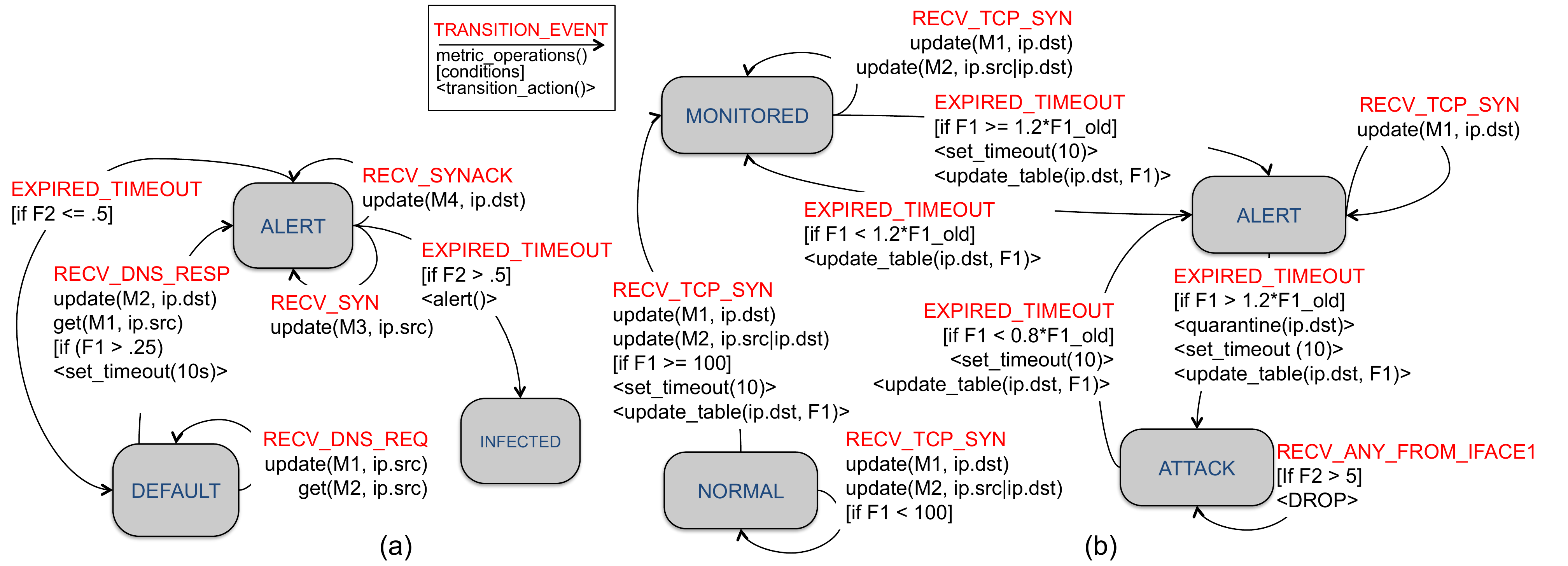}
	\caption{Application XFSMs: (a) Conficker use case; (b) DDOS use case}
	\label{fig:xfsms}
\end{figure*}

We use the following simple examples to highlight the flexibility of StreaMon in supporting heterogeneous features commonly found in real-world monitoring applications. The input data traces are obtained by properly merging a packet trace gathered from a regional Internet provider with either (i) real malicious traffic extracted from traces captured in our campus network (use case \ref{ss:use-p2p} and \ref{ss:use-conficker}) or (ii) \emph{synthetic} traces properly generated in our laboratories (use case \ref{ss:use-ddos}). 

\subsection{P2P traffic classification}
\label{ss:use-p2p}
This example shows how straightforward is the implementation of three transport layer traffic features described in \cite{kar04}, for detecting peer to peer protocols that use UDP and TCP connections. The (stateless) application considers the following packet events (which will ignore well known UDP and TCP ports.):\\
\begin{footnotesize}
$E_1: if(ip.proto==UDP)\&\&(udp.port \neq 25, 53, 110, ...)$ \\
$E_2: if(ip.proto==TCP)\&\&(tcp.port \neq 25, 53, 110, ...)$ \\
\end{footnotesize}
This application extracts the following traffic features $F_1=M_1 \& M_2$; $F_2=| M_3 - M_4|$ where:\\
$\{M_1, M_2\}$: VD enabled - return 1 for each IP src/dst pair which previously opened a \{UDP, TCP\} socket, 0 otherwise;\\
$\{M_3, M_4\}$: VD enabled and VM type CBF - count the number of different \{TCP source ports,  hosts\} connected to the same destination IP address.

Metrics are read/updated on the basis of the matched event, as follows:
\begin{center}
\begin{footnotesize}
\begin{tabular}{ | l | l | }
\hline
$E_1$ & $E_2$  \\
\hline
$set(M_1, ip.src|ip.dst)$   		& $get(M_1, ip.src|ip.dst)$; \\
$get(M_2, ip.src|ip.dst)$    	& $set(M_2, ip.src|ip.dst)$ \\
$get(M_3, ip.dst)$    			& $set(M_3, tcp.sport|ip.dst, ip.dst)$ \\
 $get(M_4, ip.dst)$       		& $set(M_4, ip.src|ip.dst, ip.dst)$ \\
				    		& $get(M_i, ip.src), i=3,4$ \\
				
\hline
\end{tabular}
\end{footnotesize}
\end{center}

The application detects a p2p client if the following condition holds after a transitory period: $(F_1==1\&\&( F_2 < 10))$. 

\subsection{Conficker botnet detection}
\label{ss:use-conficker}
Conficker is one of the largest Botnets found in recent years \cite{torpig}. A multi-step detection algorithm can attempt to track the two following (separated) phase: (1) a bot tries to contact the C\&C Server, and (2) a  single bot tries to contact and infect other hosts.

To contact the C\&C Server, infected hosts perform a huge number of DNS queries (with a high NXDomain error probability) to resolve randomly generated domains. In the \emph{infection} phase, every host tries to open a TCP connection to the ports 445 of random IPs. Our Conficker detector will use the following metrics (VD disabled and VM of type TBF ): number of total DNS queries per host ($M_1$), number of DNS NXDomain per host ($M_2$) and number of TCP SYN and SYNACK to/from port 445 (respectively $M_3$ and $M_4$). These metrics are combined into the following features:  $F_1 = M_2/M_1$, $F_2=M_4/M_3$. 

For a DNS NXDomain response, the condition $F_1 > 0.25$ is checked. If the condition is true, the state of the actual flow changes to \emph{alert} and an \emph{event timeout} is set. In the alert state the application updates $M_3$ and $M_4$ and, when this timeout expires, these metrics are used to compute $F_2$ and to verify the related condition: if the condition is true, then the host is considered \emph{infected} and goes to the next state, otherwise it returns to the default state.

\begin{figure}[t]
	\centering
	\includegraphics[width=.48\textwidth]{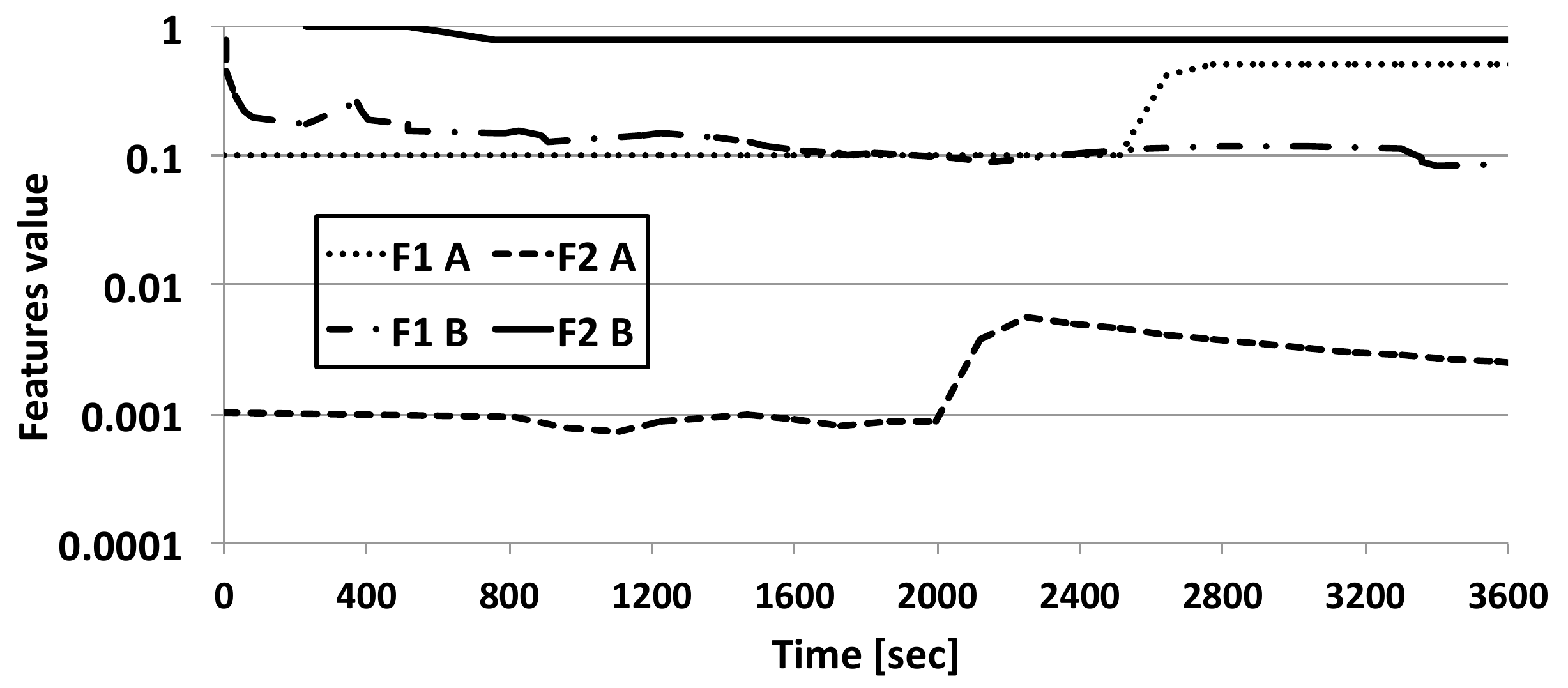}
	\caption{Conficker features temporal evolution}
	\vspace*{-.5cm}
	\label{fig:conficker}
\end{figure}

The application XFSM is graphically described (with simplified syntax) in Figure \ref{fig:xfsms}.a. Figure \ref{fig:conficker} shows the trend of features used in this configuration, for an host infected by Conficker (A) and for a \emph{clean} host (B). In the first case, the value of $F_1$ is relatively high since the beginning of the monitoring (due to the presence of reverse DNS queries, easily filtered by the application); the value increases when the host starts to perform \emph{Conficker queries}. The value of $F_2$ instead is very low, clearly denoting a port scan. Also for the host B the presence of rDNS queries increases the value of $F_1$ and this involve a change of state, and the application starts to analyze TCP feature. However the $F_2$ value (nearly 100\% of SYNACKs are received) reveals that this is clearly not a network scan. Testing this configuration in a 90-hours trace with 53 different hosts in idle state, we obtained 100\% of detection (8 infected hosts detected) without false positive or false negative.

\subsection{DDOS detection and mitigation}
\label{ss:use-ddos}
In this section we sketch a simple algorithm which can be used as an initial base for detecting and mitigating DDOS attacks. The algorithm is driven by the number of SYN packets received by possible DDOS targets.  The XFSM of this configuration is depicted in \ref{fig:xfsms}.b, and is governed by the following two events:
\\
\begin{footnotesize}
$E_1: if(ip.proto==TCP)\&\&(tcp.flags==SYN)$ \\
$E_2: timeout expired$ \\
\end{footnotesize}

Metric $M_1$ (VD=off, VM=TBF) tracks the number of TCP SYN addressed to a same target in 60 seconds (with 240s TBF's memory).  All external servers for which $F_1=M_1$ is under a given threshold (100 in this example) are in default state (because they are obviously not under attack and thus do not need an explicit status). When $F_1$ exceeds this threshold,  the target goes in \emph{monitored state}, a timeout is set and the current value of $F_1$ value is stored into a secondary support table with key ip.dst.  As soon as this timeout expires, the difference between the current feature value and the one stored is computed. If the condition $if(F_1(t) > 1.2 * F_1(t-1))$, holds for two consecutive times, i.e. the rate of TCP SYN has increased twice for more than 20\%, the flow goes into an \emph{attack} state.

Our use case example mimics this mitigation strategy by considering legitimate all the source hosts that have already shown meaningful activity and contacted the server {\em before the actual emergence of the attack}, and thus likely dropping most of the TCP SYN having a spoofed source IP address. In our use case example, a second metric $M_2$ tracks the TCP SYN rate, smoothed over a chosen time window, at which which each host contacts each destination IP addresses ($M_2$ is configured with VD disabled and VM of type TBF with smoothing window 240s, and it is updated with MFK  $(IP_{src}, IP_{dst})$.

\begin{figure}[t]
	\centering
	\includegraphics[width=.48\textwidth]{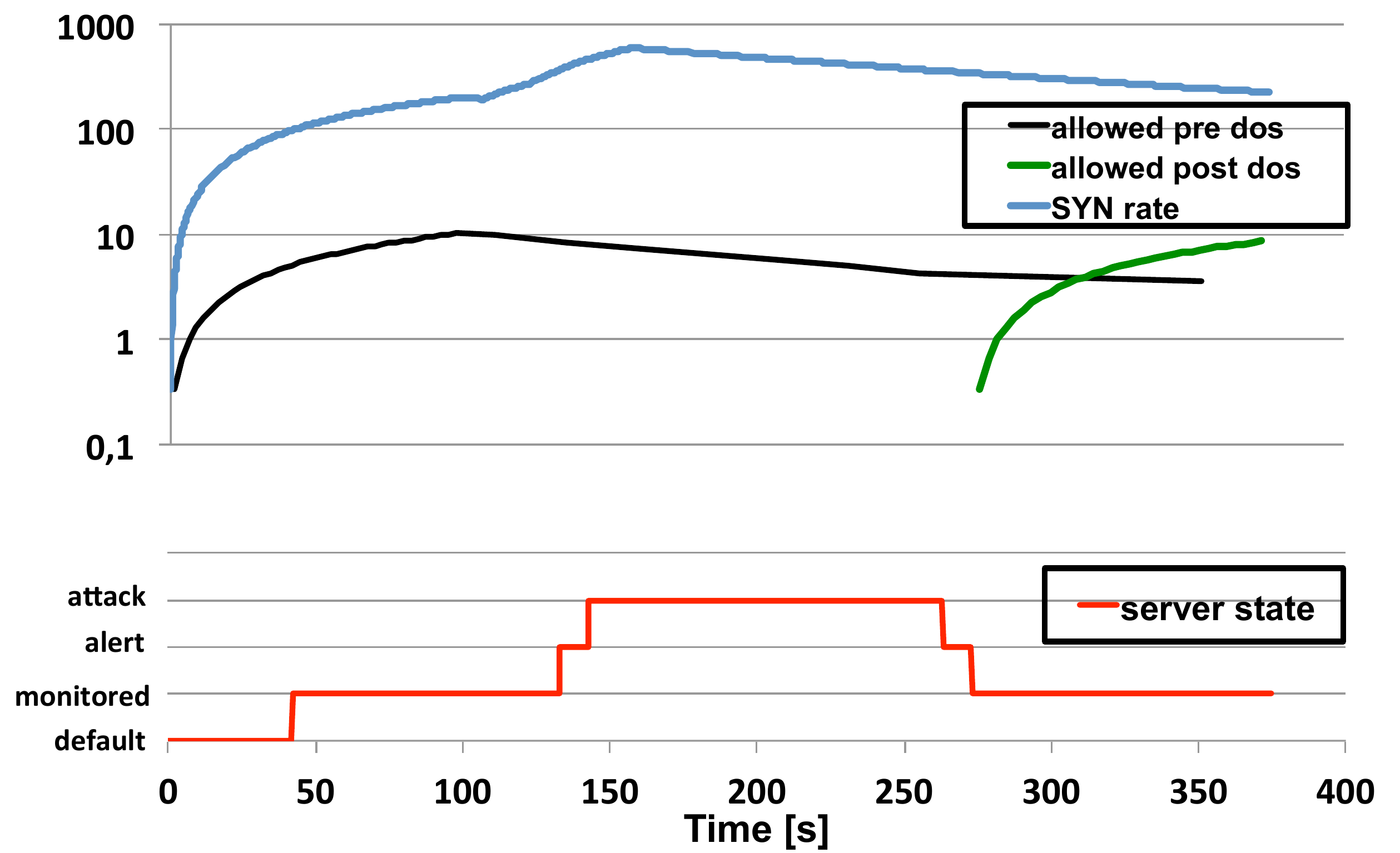}
	\caption{DDOS temporal representation: (i) DDOS target host status (red curve), (ii) $F_1$ value for the DDOS target (blu curve) and (iii) $F_2$ for two different legitimate traffic sources connecting to the DDOS target server}
	\vspace*{-.5cm}
	\label{fig:DDOS}
\end{figure}

Figure \ref{fig:DDOS} shows the temporal evolution of the state of a server, in an experiment where we performed a DDoS attack, with spoofed TCP SYN packets, after about 140s. The figure also reports the measured TCP SYN rate, as well as the traffic generated by two hosts performing regular queries towards the server: one starting at time 0, and the other starting right in the middle of the attack.
When the actual attack is detected, the mitigation strategy starts filtering traffic. Thanks to the second tracked metric, $M_2$, the user starting activities before the DDoS attack is not filtered; on the contrary, connections from sources not previously detected via the $M_2$ metric are blocked, as shown by the green curve. New connections can be accepted as soon the server leaves the \emph{attack} state (see the $F_2$ growth of the second host).

\subsection{HW accelerated detection of non standard encrypted traffic}
\label{s:entropy}
This example shows how HW metrics can be integrated in {\em StreaMon}. Since one of the task performed during deep packet inspection is the collection of statistics on byte frequencies, offsets for common byte-values, packet information entropy, we illustrate an use case that is a simplified version of the approach described in \cite{entropy}, in which encrypted flows are detected by combining two traffic features: (i) the bit information entropy of a packet; (ii) the percentage of printable characters, $i.e.$ ASCII characters in the range $[32 \ldots 127]$. Both features are computationally demanding for a software implementation, whereas they are best delegated to an FPGA implementation (just 250 logic cells for our implementation).

\begin{figure}[t]
	\centering
	\includegraphics[width=.45\textwidth]{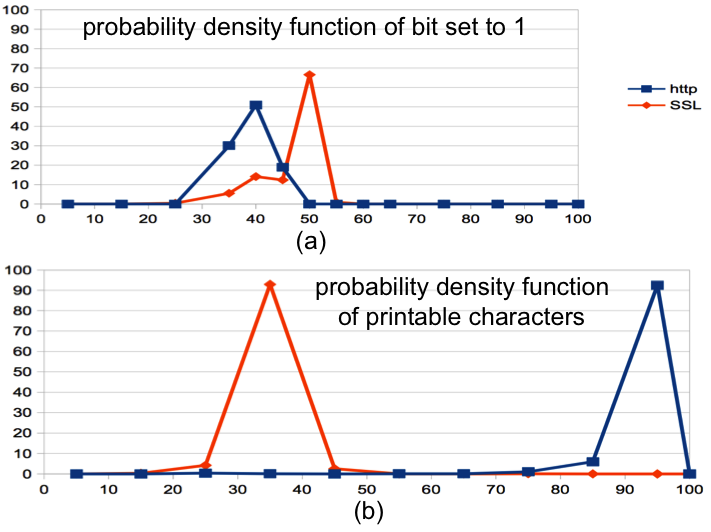}
	\caption{(a) percentage of printable character as function of the percentage of packets (b) percentage of bit set to 1 as function of the percentage of packets}
	\vspace*{-.5cm}
	\label{fig:entropy}
\end{figure}

This use case implements a simple stateless StreaMon application with the following characteristics: (i) it considers as event the reception of a UDP or TCP packet with length greater than 100 bytes; (ii) the primary key is the socket 5-tuple of the packet; (iii) it makes use of two FPGA precomputed  metrics: popcount (number of bit set to 1) $M_1$ and printable chars percentage $M_2$ (which is directly mapped into a StreaMon feature $F_1=M_1$); (v) the packet entropy Feature $F_2$ is obtained as $- \frac{1}{N} \sum_{i=0}^1 n_i \cdot (log(n_i)-log(N)$, where $N=len, n_0= N - M_1, n_1 = M_1$. An encrypted flow is detected if: $(F_2 < 0.75) \& (-3\sigma < F1- H_U(l)< 3\sigma)$

where $H_U(l)$ is the entropy of a packet with uniform distribution of 1 in the payload and length  $l$ and $\sigma$ is the standard deviation. Figure \ref{fig:entropy} shows $M_1$ and $M_2$ to the simple case of HTTP (unencrypted flow) and SSL (encrypted flow) traffic, and as in \cite{entropy} justify the condition expressed above. 

\subsection{Port knocking}
\label{s:pknock}
\begin{figure}[t]
	\centering
	\includegraphics[width=.5\textwidth]{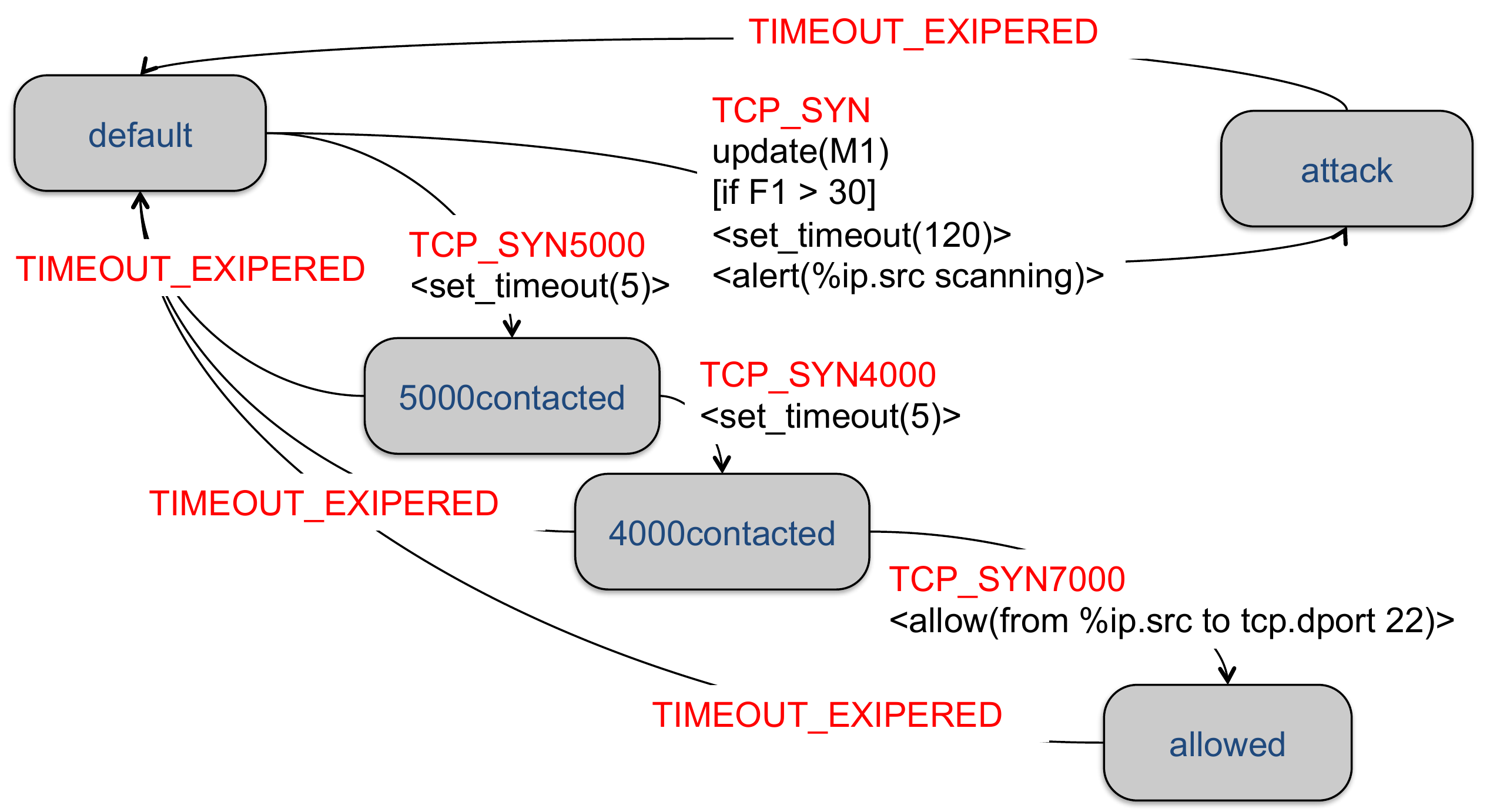}
	\vspace*{-.5cm}
	\caption{Port Knocking use case XFSM}
	\vspace*{-.5cm}
	\label{fig:portkn}
\end{figure}

This use case shows how a stateful firewall application can be easily implemented with StreaMon. In particular, this use case implements a port knocking mechanism to enable SSH access. The application's XFSM is depicted in figure \ref{fig:portkn}. SSH access will be granted only to those clients guessing the correct port sequence 5000, 4000, 7000  (a short sequence only for presentation convenience).

The application considers four events (TCP SYN received respectively on port 5000, 4000, 7000, other ports) and four states (default, 5000contacted, 4000contacted, allowed). A state transition is triggered only if the client (identified by the IP source address) contacts the expected port in the sequence. If after a state transition the next expected port is not contacted in five seconds, the flow status is rolled back to default. 

In addition, to avoid random port scanning for guessing the correct sequence, a metric $M_1$ (VD=ON, VM=TBF with smoothing window 5s) counting the SYN rate is used. If a port scan is detected ($if (F_1 = M_1) > 40$) the flow status is updated to "attack", the host is blocked for 2 minutes and an alert is generated. M1 is updated with $DFK =  ip.src|ip.dst|tcp.dport$ and $MFK = ip.src$ when a SYN to an unexpected port is contacted.


\section{Performance evaluation}


\subsection{Implementation Overview}

We experimented StreaMon on two platforms. A SW-only deployment leverages off-the-shelf NIC drivers, integrating the efficient PFQ packet capturing kernel module \cite{pfq}. To test HW acceleration, we also deployed StreaMon using a 10 gbps Combo FPGA card as packet source. In what follows, unless otherwise specified (specifically, section \ref{s:fpga} which deals with HW accelerated metrics implemented over the Combo FPGA card) we focus on the SW-only configuration.

StreaMon start-up sequence is summarized as follows. The application program is given as input to a pre-processor script as a XML formatted textual file. The program is parsed, StreaMon process is configured and executed while in parallel the interpreted feature and condition expressions are transformed them into C++ code and build a "on-demand DLL"\footnote{
	Note that this is an optimization step which is devised to on-the-fly compile (transparent to the 
	programmer), the application code, so as to avoid a "feature/condition interpreter", which would 
	have lowered the overall performance and would not have permitted line rate feature computation 
	and condition verification. Since no recompilation of the StreaMon code is ever needed, but user 
	defined features are integrated as a DLL, dynamic deployment of user programs is made possible.
}; 

\begin{figure}[!t]
\centering
   \includegraphics[width=.45\textwidth]{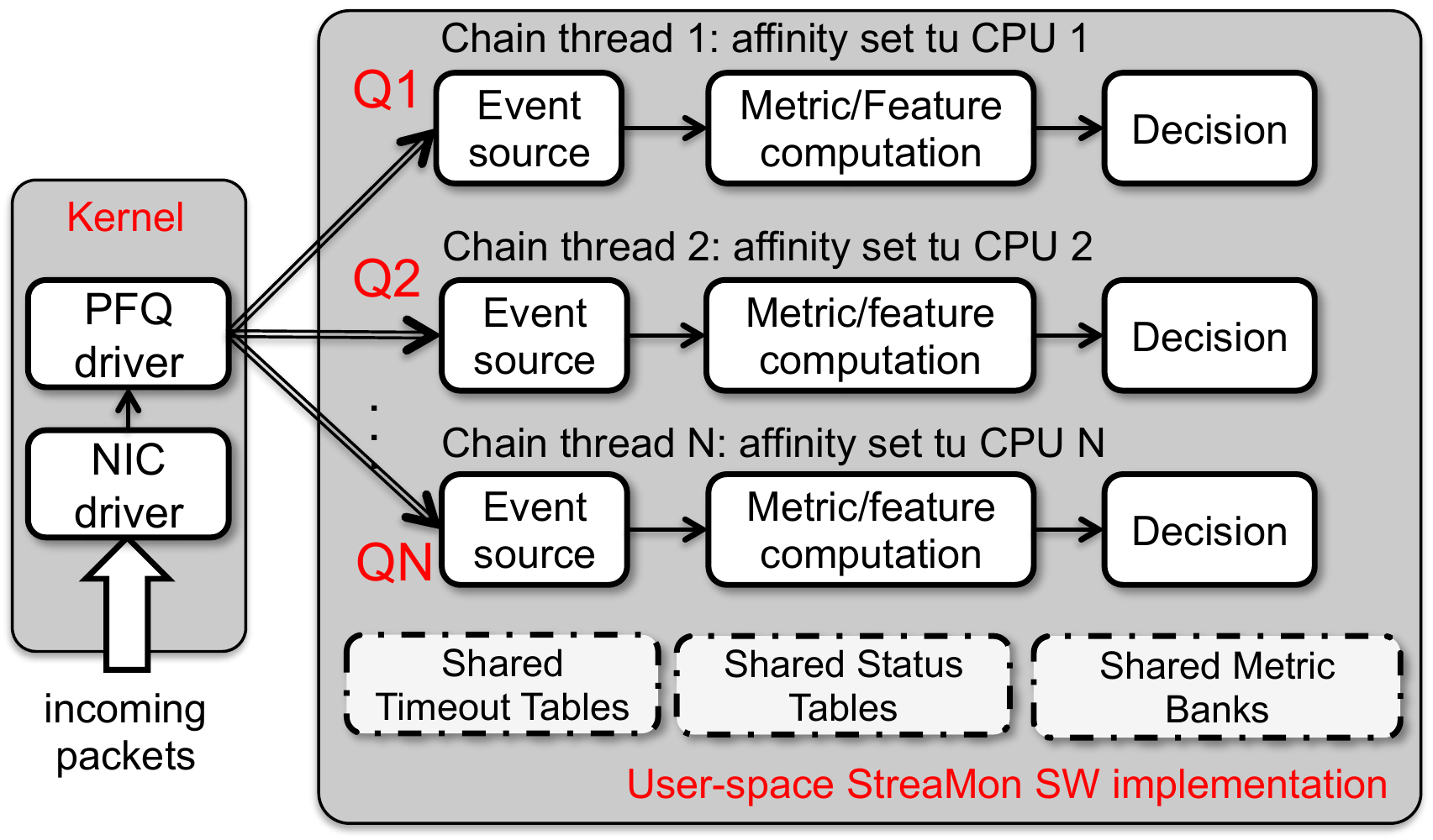}
\caption{StreaMon prototype implementation architecture}
\label{fig:multicore}
\end{figure}

StreaMon implementation takes advantages from multicore platforms by implementing parallel processing chains as shown in the implementation architecture depicted in figure \ref{fig:multicore} (for the SW-only setup). The PFQ driver is used as packet filter and configured with a number of software queues equal to the number of cores. Traffic flows are dispatched from the physical queues of the NIC by the PFQ steering function. For each PFQ, a separate StreaMon chain is executed by a single thread with CPU affinity fixed to one of the available core and share the metric banks, status tables and timeout tables (the concurrent access to the shared data is protected by spinlocks).  

The throughput measurement has been performed on a Intel Xeon X5650 (2.67 GHz, 6 cores) Linux server with, 16 GB ram and Intel 82599eb 10Gbit optical interfaces.

\subsection{Performance Analysis}
\label{s:perf}
\subsubsection{Measurement subsystem performance} 
\begin{figure}[!t]
\centering
\includegraphics[width=.45\textwidth]{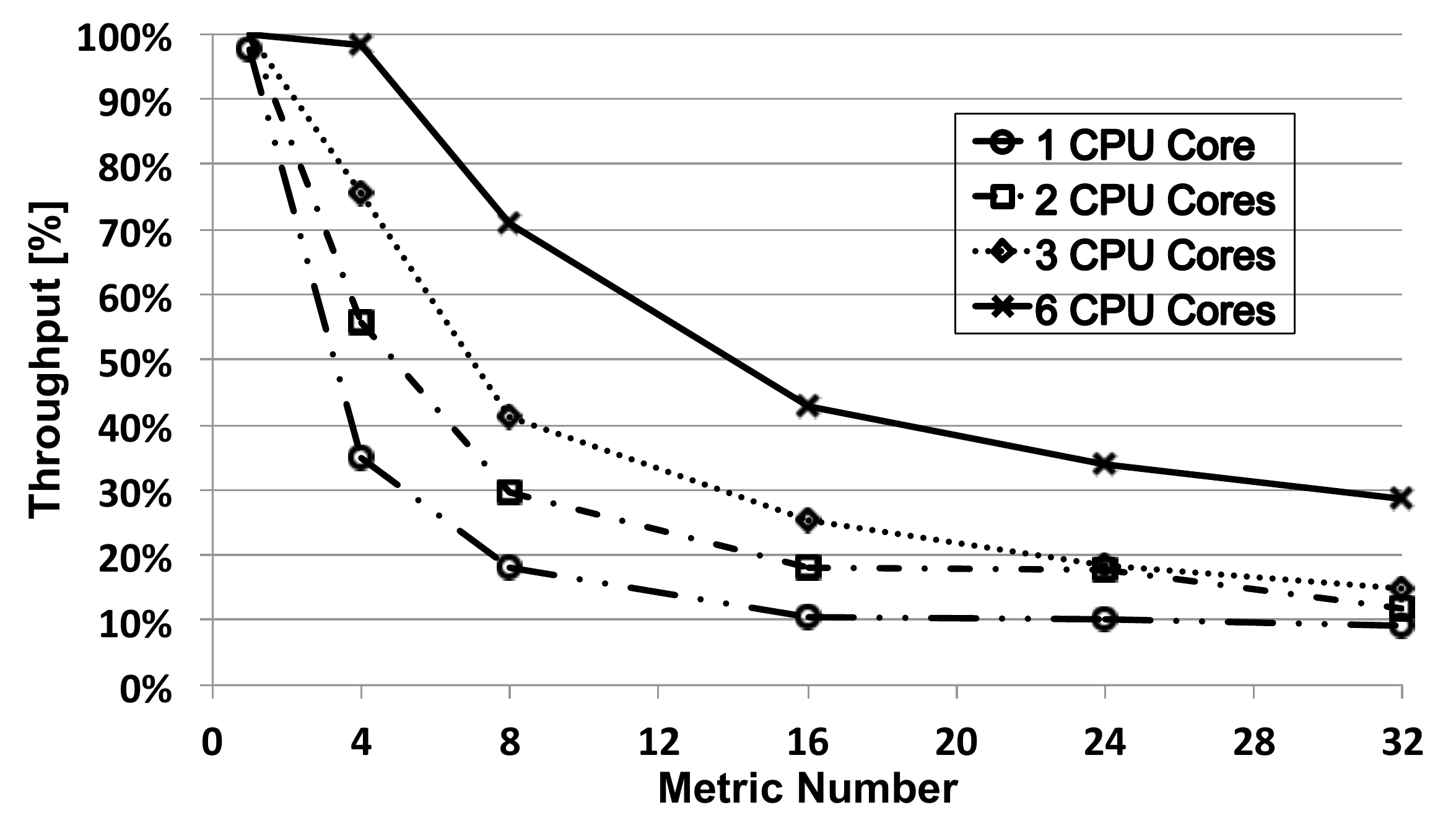}
\caption{System throughput evaluation}
\label{fig:metric-measurement}
\end{figure}

Figure \ref{fig:metric-measurement} shows the processing throughput expressed as percentage of the maximum throughput (6,468 Gbps) obtained with one PFQ source block without the overhead of StreaMon, expressed as function of the number of metrics (1, 4, 8, 16, 24, 32) in case of (1, 2, 3, 6) CPU core parallel processing. The input test data is a small portion of a long trace captured within a local internet provider infrastructure and its parameters are summarized in the following table:

\begin{center}
\begin{footnotesize}
\begin{tabular}{ | c | c | c |  c |}
\hline
\textbf{pkt no.} & \textbf{AVG pkt len} & \textbf{Host no.} & \textbf{TX rate}\\
\hline
6320928 & 632.82 bytes & 31827 & 6.47 Gbps \\
\hline
\end{tabular}
\end{footnotesize}
\end{center}

where \textit{pkt len} and \textit{TX} are the average packet length and replayed transmission bitrate respectively. 

As expected, the throughput decreases with the number of metrics and grows with the number of cores (even though the growth is lower than expected due to the simple thread concurrency management adopted in this prototype). It is important to underline that such graph shows the worst metric computational scenario in which: (i) all metrics are sequentially updated and retrieved for each packet (single event and stateless logic) and are configured with both the VD (a BF pair) and the MV (DLEFT) enabled; (ii) all flow keys are different (n metrics, n * 2 flow keys). Nevertheless, the results are promising, as for example in case of 16 metrics, for which we observe the following average bitrate (Gbps):

\begin{center}
\begin{footnotesize}
\begin{tabular}{ | c | c | c | c |}
\hline
\textbf{1 core} & \textbf{2 cores} & \textbf{3 cores} & \textbf{6 cores}\\
\hline
0.566 & 0.983 & 1.367 & 2.315 \\
\hline
\end{tabular}
\end{footnotesize}
\end{center}

Without discussing here the advantages of a stateful pre-filtering (which will be shown in section \ref{ss:use-perf}), we want to underline that, in the typical case in which not all the metrics are updated for each packet and require distinct flow keys, we can experience a better relation between the processing throughput and the number of metrics. For this reason, we compared the average throughput obtained by executing the traffic classification algorithm described in \cite{complex} (the most computational demanding algorithm cited in \ref{tab:features}, in terms of number of distinct metrics and flow keys, and portion of processed traffic). The following table shows the average bitrate (in Gbps) obtained by \cite{complex} in case of TCP traffic classification (\cite{complex}: 6 metrics, 3 flowkeys) and the average bitrate of the worst case configuration (6 metrics and 12 flow keys). 

\begin{center}
\begin{footnotesize}
\begin{tabular}{ | c | c | c | c| }
\hline
& \textbf{1 core} & \textbf{2 cores} & \textbf{3 cores} \\
\hline
\textbf{\cite{complex}} & 2.792  & 4.316 & 5.199 \\
\hline
\textbf{worst case} & 1.243 & 2.812 & 2.008 \\
\hline
\end{tabular}
\end{footnotesize}
\end{center}

\begin{table*}
\hfill{}
\begin{center}
\begin{footnotesize}
\begin{tabular}{ | c | c | c | c | c | l |}
\hline
\textbf{case} & \textbf{pkts}. & \textbf{AVG plen [bytes]} & \textbf{Ms, FKs.}& \textbf{TX [Gbps]} & \textbf{RX [\# core: Gbps]}\\
\hline
\ref{ss:use-p2p} & 7676958 & 404.97 & 8, 4  & 3.658 & 1: 1.956,  2: 2.954,  3: 3.654\\
\hline
\ref{ss:use-conficker} & 11852112 & 329.77 & 4, 2 & 2.462 & 1: 2.462 \\
\hline
\ref{ss:use-ddos} & 9387240 & 215.38 & 2, 2 & 2.067 & 1: 2.003, 2: 2.067\\
\hline
\end{tabular}
\caption{Use case trace parameters and throughput}
\vspace*{-.5cm}
\label{tab:usecases}
\end{footnotesize}
\end{center}
\end{table*}

\subsubsection{Use case performance evaluation}
\label{ss:use-perf}
Table \ref{tab:usecases} shows the details of the test traces used for three among the use cases presented in Section \ref{ss:usecases} and the their performance evaluation, where \emph{pkt. no} is the total number of packets, \emph{AVG plen} is the average packet length, \emph{M} is the total number of metrics, \emph{FK} is that the transmission bitrate is fixed by the injector probe and depends on the average packet length. 

As already underlined, the use case applications experience a better throughput with respect to the bitrate obtained with the same number of metrics in the worst case measurement showed in \ref{fig:metric-measurement}. Note that due to the different performances of the COMBO card processor and the limitation to single core processing\footnote{This is due the fact that the COMBO card capture driver does not allow to open multiple instances of the same communication channel toward the FPGA NIC}, the encrypted flow detection use case performance is omitted to avoid confusion, as it would present results not directly comparable with the remaining ones. 

\subsubsection{FPGA accelerated primitives}
\label{s:fpga}
As already discussed, our current StreaMon prototype supports seamless integration of HW precomputed 
primitives, and in particular we have implemented packet entropy computation (see section \ref{s:entropy}) and event parsing on a INVEA-TECH Combo FPGA card \cite{invea}. The interface between the Combo card and StreaMon core software is realized via a footer appended to the ethernet frame. The Combo card capture the packet traveling in the network under inspection, performs the required operation and transfer the resulting metrics and signals as a list of Type Len Value (TLV) to the PC hosting the capture card. This list of TLV is parsed by the Event Layer and will be available to the remaining layers. 

\begin{figure}[!t]
\centering
\includegraphics[width=.4\textwidth]{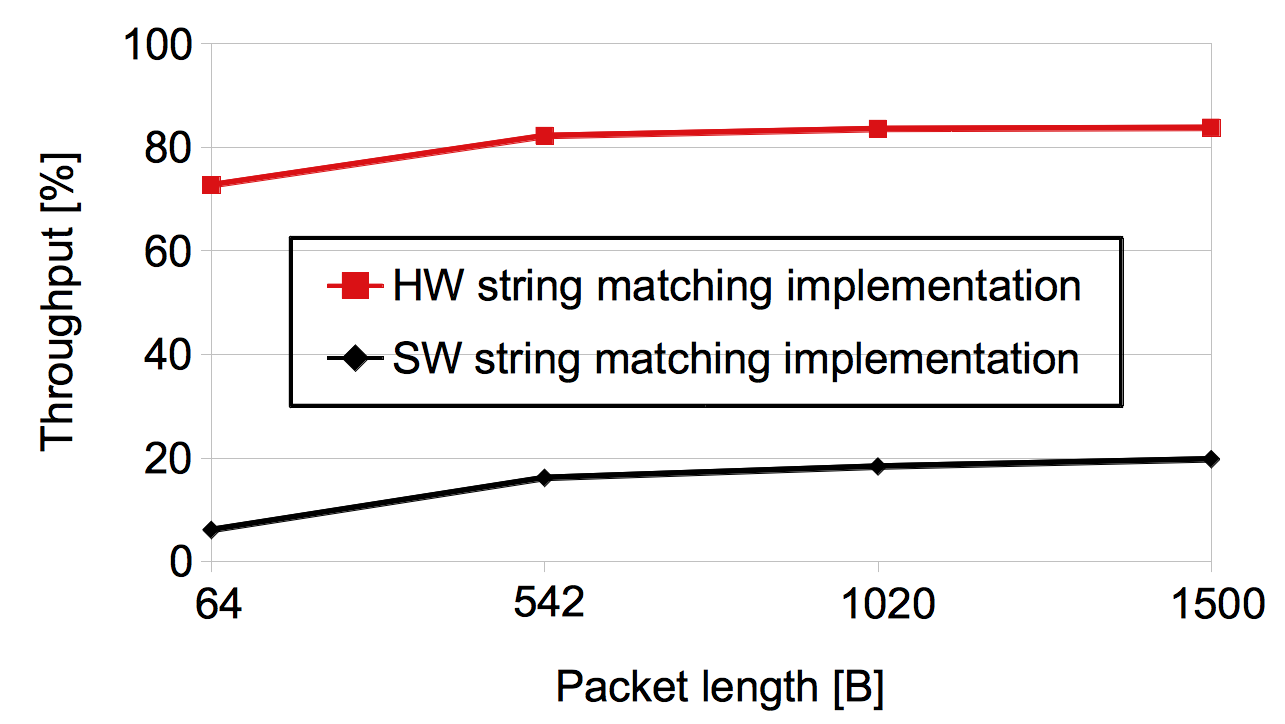}
\caption{FPGA accelerated and SW string matching comparison}
\vspace*{-.5cm}
\label{fig:string_matching}
\end{figure}

Our next step is to implement a full deep packet inspection module. To confirm its necessity, we have so far implemented a simplified content matching packet payload inspection primitive in both SW and HW. Figure \ref{fig:string_matching} compares the two implementations for a toy example of just 32 content strings, showing the severe performance degradation of the SW implementation even in such small scale example. At the same time,  
the figure shows that, by offloading content matching primitives to the HW front-end, our prototype achieves almost optimal throughput performance. 

\section{Related Work}
In the literature, several monitoring platforms have targeted monitoring applications' programmability. 
A Monitoring API for programmable HW network adapters is proposed in \cite{trimintzios06dimapi}. 
On top of such probe, network administrators may implement custom C++ monitoring applications. 
One of the developed applications is Appmon \cite{antoniades06appmon}. It uses deep packet 
inspection to classify observed flows and attribute these flows to an application. Flow \emph{states} 
are stored in an hash table and retrieved when an \emph{old} flow is observed again. This way to 
handle states bears some resemblance with that proposed in this work, which however makes usage 
of (much) more descriptive eXtended Finite State Machines. CoMo \cite{ian04} is another well known 
network monitoring platform. We share with CoMo the (for us, side) idea of extensible plug-in metric 
modules, but besides this we are quite orthogonal to such work, as we rather focus on how to combine 
metrics with features and states using higher level programming techniques (versus CoMo's low level queries). 

Bro \cite{bro} provides a monitoring framework relying on event-based programming language for real time
 statistics and notification. Despite the attempt to define a versatile high level language, Bro is not designed 
 to expose a clear and simple abstraction for monitoring application development and leave full programmability
  to its users (which we believe results in a more descriptive and yet more complex programming language).

The Real-Time Communications Monitoring (RTCMon) framework \cite{rtcmon} permits development of 
monitoring applications, but again the development language is a low level one (C++), and (unlike us) 
any feature extraction and state handling must be dealt with inside the custom application logic developed 
by the programmer.  CoralReef \cite{keys01coralarchitecture}, FLAME \cite{flame02} and Blockmon \cite{Block12} 
are other frameworks which grant full programmability by permitting the monitoring application developers to 
``hook'' their custom C/C++/Perl traffic analysis function to the platform. 
On a different line, a number of monitoring frameworks are based on suitable extensions of Data Stream 
Management Systems (DSMS). PaQueT \cite{lig08paquet}, and more recently BackStreamDB \cite{Back12}, 
are programmable monitoring frameworks developed as an extension of the Borealis DSMS \cite{borealis05}. 
Ease of programming and high flexibility is provided by permitting users to define new \emph{traffic} metrics 
by simply performing queries to the DSMS. The DSMS is configured through an XML file that is processed to 
obtain a C++ application code. Gigascope \cite{cranor03gigascope} is another stream database for network 
monitoring that provide an architecture programmable via SQL-like queries.

Opensketch \cite{opensk} is a recent work proposing an efficient generic data plane based on programmable 
metric sketches. If on the one hand we share with Opensketch the same measurement approach, on the other 
hand its data plane abstraction delegates any decision stage and logic adaptation the control plane and, with 
reference to our proposed abstraction, does not go beyond the functionalities of our proposed Measurement 
Subsystem. On the same line, ProgME \cite{progme} is a programmable measurement framework which revolves 
around the extended and more scalable notion of dynamic flowset composition, for which it provides 
a novel functional language. 

Even though equivalent dynamic tracking strategies might be deployed over Openflow based monitoring tools, 
by exploiting multiple tables, metadata and by delegating "monitoring intelligence"  to external controllers, this 
approach would require to fully develop the specific application logic and to forward all packets to an external 
controller, (like in Openflow based monitoring tool Fresco \cite{fresco}), which will increase complexity 
and affect performance.

Finally, while our work is, to the best of our knowledge, the first which exploits eXtended Finite State Machines 
(XFSM) for programming custom monitoring logic, we acknowledge that the idea of using XFSM as programming language for networking purposes was proposed in a completely different field (wireless MAC protocols programmability) by \cite{bia12}.

\section{Conclusion}
\label{s:conclusion}
The StreaMon programmable monitoring framework described in this paper aims at making the deployment of monitoring applications as fast an easy as configuring a set of pre-established metrics and devising a state machine which orchestrates their operation while following the evolution of attacks and anomalies. Indeed, the major contribution of the paper is the design of a pragmatic application programming interface for developing stream-based monitoring tasks, which does not require programmers to access the monitoring device internals.  Despite their simplicity, we believe that the wide range of features accounted in the proposed use cases suggest that StreaMon's flexibility can be exploited to develop and deploy several real world applications. 




\footnotesize
\bibliographystyle{abbrv}
\bibliography{submit}

\end{document}